\documentclass[twocolumn,showpacs,preprintnumbers,amsmath,amssymb]{revtex4}

\usepackage{graphicx}
\usepackage{dcolumn}
\usepackage{bm}
\usepackage{float}

\begin{document}

\title{Arbitrary spin in a spin bath: Exact dynamics and approximation techniques}

\author{Vitalii Semin}
 \email{semin@ukzn.ac.za}
\affiliation{Quantum Research Group, School of Chemistry and Physics,
 University of KwaZulu-Natal, Durban, 4001, South Africa}

\author{Ilya Sinayskiy}
 \email{sinayskiy@ukzn.ac.za}
\affiliation{
Quantum Research Group, School of Chemistry and Physics,  University of KwaZulu-Natal, Durban, 4001, South Africa and National Institute for Theoretical Physics (NITheP), KwaZulu-Natal, South Africa}

\author{Francesco Petruccione}
 \email{petruccione@ukzn.ac.za}
\affiliation{
Quantum Research Group, School of Chemistry and Physics,  University of KwaZulu-Natal, Durban, 4001, South Africa and National Institute for Theoretical Physics (NITheP), KwaZulu-Natal, South Africa}

\date{\today}

\begin{abstract}

A model of an arbitrary spin coupled to a bath of spins 1/2 in a star configuration is considered. The exact reduced dynamics of the central spin is found for the case of non-correlated initial conditions of the system and the bath. The exact solution is used 
to test two approximation techniques, namely, the Nakajima-Zwanzig 
projection operator technique and the time-convolutionless 
projection operator technique corresponding to the second order of the coupling constant. Two types of  projection operators are used for deriving the master equations and the results are compared with the exact solution for a central spin equal to one. It is shown that the approximate master equations reproduce the exact dynamics on time-scales $1/(A\sqrt{N})$, where $A$ is the coupling constant and $N$ is the number of spins in the bath.
\end{abstract}


\maketitle

\section{INTRODUCTION}
Many problems in quantum physics can be considered using a model in which one or
more mesoscopic or even macroscopic coordinates interact with a background
environment \cite{1}. Most environments can be modelled as  a set of oscillators or  as a set of spins 1/2 \cite{SPIN}. The physical models in which quantum systems interact with a spin bath play a role in the quantum theory of magnetism \cite{itqss}, quantum spin glasses \cite{glas}, theory of conductors \cite{cond,cond2} and superconductors \cite{super}. However, typical models of this type are usually complicated and cannot be described exactly \cite{toqs}.

In order to describe the dynamics of a typical spin-bath model one needs to use appropriate approximation techniques \cite{toqs}. All approximation techniques can be divided into two classes: local in time and non-local in time approaches. The most widely used non-local in time approach is the Nakajima-Zwanzig projection operator technique \cite{ZWANZIG, NAKAJIMA}. The typical example of the local in time technique is a time-convolutionless  projection operator technique \cite{TCL_first,toqs}. For the same order of the perturbation expansion both approaches accurately describe the non-Markovian dynamics of the reduced system and  give similar results for the timescales of the order $1/\gamma$, where $\gamma$ is a spontaneous emission  constant \cite{oldTCL, CRL}. The crucial point in both approaches is the appropriate choice of a projection operator, which separates the relevant part of the total density matrix from the irrelevant one \cite{Repke}. A successful choice of a projection operator can lead to a simpler form of the reduced master equation and to higher accuracy of the approximation technique.

In this article we study a model of an arbitrary spin resonantly coupled to a spin bath in a star configuration. Due to the special choice of the uniform coupling of the central spin to the bath spins, it is possible to find an exact evolution operator and to construct an explicit form of the reduced dynamics of the central spin. This result generalize the results presented in \cite{4,CRL,fannes} for the cases of central spins equal to 1/2 and 1. The aim of this work is to test the most commonly used local and non-local in time approximation techniques for different choices of projection operators. We construct  the second order Nakajima-Zwanzig and the second order time-convolutionless master equations for traditional and correlated projection operators \cite{CRL}.

This article is organised as follows. In Sec. II we describe the
model of an arbitrary  spin coupled to a spin bath, present the analytical solution for
the evolution operator and build the reduced density matrix for the central spin. In Sec. III we apply different projection operator methods to obtain two types of quantum master equations, and the general theory is used for a model spin equal to 1 coupled to a spin bath.  Finally, in Sec. IV we discuss the results and conclude. The analysis of the non-resonant case of a central spin equal to 1/2 in a spin bath is presented in appendix.

\section{MODEL AND ITS EXACT DYNAMICS}
We consider a model of an arbitrary spin coupled to a bath of $N$ 1/2-spins. The model Hamiltonian  is given by
\begin{equation}\label{ham_in}
H=\sum\limits_{k=1}^{N}A_k\left(S_x\sigma_x^k+S_y\sigma_y^k+S_z\sigma_z^k \right), 
\end{equation}
where $\sigma_x^k$, $\sigma_y^k$, and $\sigma_z^k$ are the Pauli matrices of the $k$-th spin in the bath, $S_i$ is the component of the relevant spin along the $i$-axis, $N$ is the number of spins in the bath, and $A_k$ are  the strengths of the interaction between the central spin and the $k$th spin of the bath. 
 We take the uniform coupling of the central spin and the spin bath to be $A_k=A$. The model can then be solved analytically as follows.

The uniform coupling Hamiltonian \eqref{ham_in} can be rewritten in the form
\begin{equation}
H=A\left(S_x J_x+S_y J_y+S_z J_z \right)=A\mathbf{SJ}. \label{hen_ham}
\end{equation} 
In the above equation, $J_i=\sum\limits_{k}^{N}\sigma_i^k$ are the collective operators of the bath spins, $\mathbf{S}$ and $\mathbf{J}$ are the vectors of the full spin of the central spin and the bath, respectively. 

The expression \eqref{hen_ham} can be written as a sum of commutating terms, namely
\begin{equation}
H=A\mathbf{SJ}=\frac{A}{2}\left(M^2-S^2-J^2\right), \label{hen_main}
\end{equation} 
where $\mathbf{M}=\mathbf{J}+\mathbf{S}$ is a vector of the total spin of the total system. In this paper units are chosen such that $k_B=\hbar=1$, where $k_B$ is Boltzmann's constant and $\hbar$ is Plank's constant. 

To describe the exact dynamics of the total system we need to specify an initial state of the total system which is given by the  density operator $\rho_{\mathrm{tot}}(0)$ and to find an evolution operator of the total system $U(t)$ in an explicit form, as
\begin{equation}U(t)=\exp[-i H t]\label{U}.\end{equation} 
With the knowledge of the evolution operator and the initial state of the total system the reduced dynamics of the central  spin can be found as  
\begin{equation}\label{reduced}
\rho_{S}(t)=\mathrm{tr}_{B}\{U(t)\rho_{\mathrm{tot}}(0)U^{\dagger}(t)\}.
\end{equation} 
Hereafter we will consider an initially uncorrelated state between the relevant spin and the bath. The initial state of the total system reads,
\begin{equation}\label{initial}
\rho_{\mathrm{tot}}(0)=\rho_{S}(0)\otimes \rho_{B}(0),
\end{equation} 
where $\rho_{S}(0)$ is the density matrix of the relevant spin, while  $\rho_B(0)$ is the density matrix describing the state of $N$ particle with  1/2-spins, as
\begin{equation}
\rho_{B}(0)=\frac{\exp[-\beta J_z]}{Z}.
\end{equation} 
In the above expression $\beta=1/T,$ where $T$ is the temperature and $Z=\mathrm{tr}\exp[-\beta J_z]$

From Eqs. ~\eqref{U} and \eqref{hen_main} one easily finds 
\begin{equation}\label{sol}
U=\exp[- i \frac{A}{2}t M^2]\exp[ i \frac{A}{2}t S^2]\exp[ i \frac{A}{2}t J^2].
\end{equation} 
Combining Eqs.~\eqref{reduced}, \eqref{initial} and \eqref{sol} and using the properties of the trace one finds
\begin{eqnarray}\label{solgen}
\rho_S(t)=\frac{1}{Z}\mathrm{tr}_{B}\exp[-i A t M^2/2]\left\lbrace \rho_{S}(0)\right.  \\ \left. \otimes \exp[-\beta J_z]\right\rbrace \exp[i A t M^2/2]\nonumber.
\end{eqnarray}
In the construction of the above equation we used the fact that the operators $S^2$ commute with any operator in a given representation. The partition function $Z$ in Eq.~\eqref{solgen} is
 \begin{equation}\label{z}
 Z=2^N \cosh^{N}(\beta/2).
 \end{equation}
 
 Finally,  one finds the exact dynamics of a matrix element of the reduced density matrix in the basis of the eigenvectors of the operators $S^2$ and $S_z$ as
\begin{widetext}
\begin{eqnarray}\label{exact}
\langle j_1 m|\rho_S(t)|j_1 \tilde{m}\rangle=\sum\limits_{j_2}^{N/2}\frac{N_{j_2}}{z}\sum\limits_{J=|j_1-j_2|}^{j_1+j_2}\sum\limits_{J'=|j_1-j_2|}^{j_1+j_2}
\sum\limits_{\Delta=-j_1-\mathrm{min}\{m,\tilde{m}\}}^{j_1-\mathrm{min}\{m,\tilde{m}\}}\sum\limits_{m_2=-j_2}^{j_2}e^{-i A t(J(J+1)-J'(J'+1))/2-\beta m_2}\\
\times \langle j_1 m-\Delta|\rho_S(0)|j_1 \tilde{m}-\Delta\rangle C_{j_1\tilde{m}-\Delta j_2 m_2}^{J' \tilde{m}+m_2-\Delta}C_{j_1 m-\Delta j_2 m_2}^{J m+m_2-\Delta}
C_{j_1 m j_2 m_2-\Delta}^{J \tilde{m}+m_2-\Delta}C_{j_1\tilde{m} j_2 m_2-\Delta}^{J' \tilde{m}+m_2-\Delta}\nonumber,
\end{eqnarray}
\end{widetext}
where $C^{JM}_{j_1m_1j_2m_2}$ are the known Clebsch-Gordon coefficients \cite{Varsh} and $N_j=
 \binom{N}{\frac{N}{2}+j}-\binom{N}{\frac{N}{2}+j+1}$ is the degeneracy  \cite{4}. 
 The sum over $j_2$ runs from $0$   ($N$ even)  or $1/2$ ($N$ odd) to $N/2$. 
 
Particular cases of the exact solution (Eq.~\eqref{exact}) 
 are presented in the Figs.~1, 2 and 3.  The number of spins in the bath has a weak influence on the dynamics of the central spin (see Fig.~1) and for a large number of spins in the bath $N>200$ the dynamics  of the central spin is almost independent from the number of spins in the bath. The influence of the inverse temperature (see Fig.~2) consists in an increase of the probability to find the spin in the exited state with increasing of $\beta$. Fig.~3 shows the probability of the filling of the level $|j,1\rangle$ for different values of $j$. The growth of the value of a spin corresponds to an increase of the degree of freedom in the system. So, the probability to find the central spin in some state decreases with increasing  of $j$.
 
 The analysis of the exact solution shows the periodic nature of the behaviour of the central spin. The period of the solution equals to $2\pi/A$ if the total spin of the system is integer and $4\pi/A$ if the total spin of the system is half-integer. Moreover, the dynamics of the central spin is unitary and time reversible. This conclusion holds for any value of the central spin. The above result generalizes the results presented in \cite{4} and \cite{CRL} for the case of the central spin equal to 1/2 and results \cite{fannes} for the central spin equal to 1.

 \begin{figure}
 \includegraphics[scale=0.5]{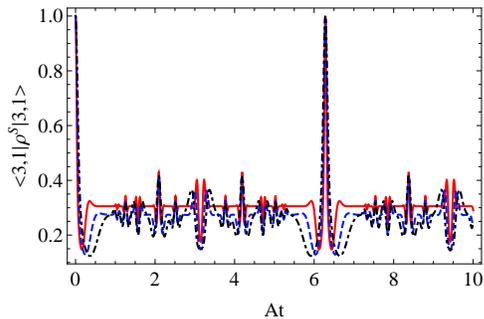}
 \caption{(Color online) The exact dynamics of the state $|3,1\rangle$ for the spin $j=3$ in a spin bath for the different numbers $N$ of spins in the bath ($N=200$ (red solid curve), $N=100$ (blue dashed curve) and $N=50$ (black dot-dashed curve)). The inverse  bath temperature is $\beta=1/T=0.25.$ The initial state is $\rho_S(0)=|3,1\rangle\langle3,1|$. The coupling constant is $A=1.$} \label{fig1}
 \end{figure}
 
 \begin{figure}
 \includegraphics[scale=0.5]{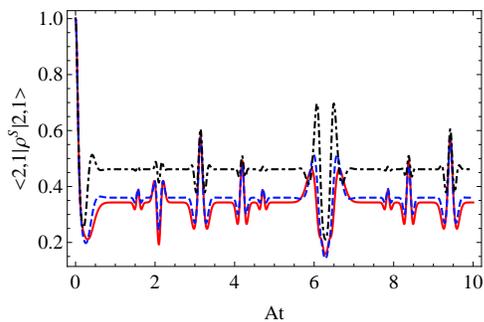}
 \caption{(Color online) The exact dynamics of the state $|2,1\rangle$ for the spin $j=2$ in the spin bath for the different inverse temperatures $\beta$ of the bath 
  ($\beta=0$ (red solid curve),  $\beta=0.25$ (blue dashed curve), $\beta=0.5$ (black dot-dashed curve)). The number of spins in the bath is $N=101$. The initial state is $\rho_S(0)=|2,1\rangle\langle2,1|$. The coupling constant is $A=1.$}\label{fig2}
 \end{figure}

 \begin{figure}
 \includegraphics[scale=0.5]{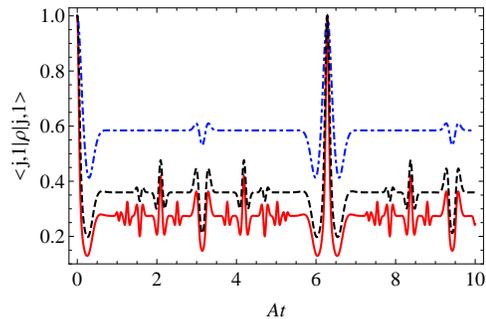}
 \caption{(Color online) The exact dynamics of the central spin in a spin bath in state $|j,1\rangle$. The value of the central spin is: $j=3$ (red solid curve), $j=2$ (black dashed curve) and $j=1$ (blue dot-dashed curve). The inverse bath temperature is $\beta=1/T=0.25.$ The initial state is $\rho_S(0)=|j,1\rangle\langle j,1|$. The number of spins in the bath $N=100$. The coupling constant is $A=1.$} 
 \end{figure}

\section{APPROXIMATION TECHNIQUES}
In this section we will apply approximation techniques to the model discussed in the previous section. Due to 
the presence of the exact reduced dynamics of this model we can examine various approximate master equations for the density matrix of the central spin and analyse their accuracy in leading order perturbation expansions. 

For convenience of the further investigation, we perform a unitary transformation of the Hamiltonian \eqref{hen_ham} 

\begin{eqnarray}\label{intham}
H_I=\frac{A}{2}e^{i A S_zJ_zt}\left( S_-J_++S_+J_-\right) e^{-i A S_zJ_zt}\\
=\frac{A}{2}\left(S_+e^{-i A(S_z-J_z)t}J_-+\mathrm{H.c.}\right). \nonumber
\end{eqnarray}
Because the exact result \eqref{exact} was found in another picture, we must transform back to the old picture all further results to be able to compare to the exact ones as follows

\begin{eqnarray}\label{backtrans}
O=e^{-i A S_zJ_zt}O_I e^{i A S_zJ_zt},
\end{eqnarray}
where $O$ and $O_I$ are arbitrary operators in the old and the new pictures, respectively. The corresponding transformation for the density matrix reads

\begin{eqnarray}\label{backtransr}
\rho(t)=e^{i A S_zJ_zt}\rho_I(t) e^{-i A S_zJ_zt}.
\end{eqnarray}

\subsection{Second-order approximations}

The general form of the second order Nakajima-Zwanzig master equation for the relevant part of the density operator for the class of initial conditions \eqref{initial} can be written as \cite{NAKAJIMA,ZWANZIG} 
\begin{equation}\label{NJgen}
\frac{\partial}{\partial t}\mathcal{P}\rho=-\int\limits_{0}^{t}dt_1\mathcal{P}[H_I(t),[H_I(t_1),\mathcal{P}\rho(t_1)]].
\end{equation}
In the above equation $H_I$ is the Hamiltonian \eqref{intham} and $\mathcal{P}$ is some projection operator. 
The second order Nakajima-Zwanzig master equation \eqref{NJgen} is closely related to a second order time-convolutionless master (TCL) equation \cite{TCL_first}. The second order TCL master equation can be derived from Eq.~\eqref{NJgen} replacing $\mathcal{P}\rho(t_1)$ by $\mathcal{P}\rho(t).$ Explicitly, the second order TCL master equation is given by
\begin{equation}\label{TCLgen}
\frac{\partial}{\partial t}\mathcal{P}\rho=-\int\limits_{0}^{t}dt_1\mathcal{P}[H_I(t),[H_I(t_1),\mathcal{P}\rho(t)]].
\end{equation}
The characteristic feature of the TCL master equation is its locality in  time.
Nevertheless, the differential equation \eqref{TCLgen} is equivalent to the integro-differential equation \eqref{NJgen} and, thus, describes memory effects. Both types of master equations are build from general assumptions, have the same general structure for a broad class of models,  and the choice of a projection operator is almost arbitrary. A review of the most important classes of  projection operators can be found in  \cite{Repke}. In the following section we will describe few possible types of projection operators.

\subsection{Projection operators}
A general class of time-independent projection operator was build in \cite{CRL,CRL0,CRL2}. Any projection operator can be represented as follows,
\begin{equation} \label{PROJECTION-GENFORM}
 {\mathcal{P}}\rho = \sum_i {\mathrm{tr}}_E \{ A_i \rho \}
 \otimes B_i,
\end{equation}
where $\{A_i\}$ and $\{B_i\}$ are two sets of linear independent
Hermitian operators on the Hilbert space of the environment of the open system ${\mathcal{H}}_B$ satisfying the relations
\begin{eqnarray}
 {\mathrm{tr}}_B \{ B_i A_j \} &=& \delta_{ij}, \label{BjAi} \\
 \sum_i ({\mathrm{tr}}_B B_i) A_i &=& I_E, \label{TRACE-PRESERVING} \\
 \sum_i A_i^T \otimes B_i &\geq& 0. \label{COND-POS}
\end{eqnarray}
Once $\mathcal{P}$ is chosen, the dynamics of the open system is
uniquely determined by the dynamical variables
\begin{equation}
    \rho_i(t) = \mathrm{tr}_B \{ A_i \rho(t) \}.
\end{equation}
The connection to the reduced density matrix is simply given by
\begin{equation}\label{rhosum}
    \rho_S(t) = \sum_i \rho_i(t),
\end{equation}
and the normalization condition reads
\begin{equation}
    \mathrm{tr}_S \> \rho_S(t) = \sum_i \mathrm{tr}_S \> \rho_i(t) = 1.
\end{equation}
If one fixes some set of operators with the properties \eqref{BjAi}-\eqref{COND-POS}, then the projection operator \eqref{PROJECTION-GENFORM} is fully defined and the dynamics of the system under investigation is governed by one of the master equations presented above.

The most simple choice of a projection  operator is an operator that projects on some fixed state of the environment, for instance, on an equilibrium state. In this case $A_n=I$ and $B_i=\exp(-\beta H_E)/Z$, where $H_E$ is a free Hamiltonian of the environment and $I$ is the identity operator. The projection operator \eqref{PROJECTION-GENFORM} then is
\begin{equation} \label{SimpleProj}
 {\mathcal{P}}\rho ={\mathrm{tr}}_E \{\rho \}
 \otimes \exp(-\beta H_E)/Z.
\end{equation}
Notice that for infinite temperature the projection operator simplifies to ${\mathcal{P}}\rho ={\mathrm{tr}}_E \{\rho \}
 \otimes I/Z$. In this form the projection operator \eqref{SimpleProj} applies to most open quantum systems.

Another way to build the projection operator \eqref{PROJECTION-GENFORM} was described in  \cite{CRL,CRL0,CRL2}.  It was proposed  to choose operators in \eqref{PROJECTION-GENFORM} from a class of projectors, which project on a subspace of some preserved quantities of the system of interest. Example of conserved quantities of open quantum 
system can be, e.g., the total number of particles or the total energy of the system.

\subsection{Application to the model}
In this subsection we apply the above approximation techniques to the model of an arbitrary spin coupled to a spin bath, as discussed in Sec. II. The first projection operator  \eqref{SimpleProj} has the form 

\begin{equation}\label{1po}
\rho_1^S=\mathcal{P}_1\rho=(\mathrm{tr}_B \rho)\otimes e^{-\beta J_z}/Z,
\end{equation}
where $Z$ is defined in Eq.~\eqref{z}. The inverse transformation \eqref{backtrans} for the density matrix \eqref{1po} is defined by
\begin{equation}
\rho_1^S=\sum\limits_{j=0,1/2}^{N_2/2}\sum\limits_{m=-j}^{j}e^{iA m S_z}\rho_1^Se^{-iA m S_z}.
\end{equation}
It is clear, that the transformation does not change the diagonal elements.

We notice \cite{my}, that the spin system has two naturally conserved quantities, namely, the projection of the total spin on the $z$-axis $S_z$ and the square of the full spin $S^2$. So we can build another projection operator for the underlying model of the form

\begin{eqnarray} \label{2po}
\mathcal{P}_2\rho&=&\sum\limits_{j=0,\frac{1}{2}}^{N/2}\sum\limits_{m=-j}^{j}\mathrm{tr}_{B}\left(\Pi_{jm}\rho\right)\otimes \frac{\Pi_{jm}}{N_j}, 
\end{eqnarray}
where $\Pi_{jm}=|jm\rangle\langle jm|$ is the projection operator on the eigenvectors of the bath operator $J_z$ and $J^2$ and $N_j =\mathrm{tr}\Pi_{jm}=
 \binom{N}{\frac{N}{2}+j}-\binom{N}{\frac{N}{2}+j+1}.$ The  operator \eqref{2po} fulfils the properties \eqref{BjAi}-\eqref{COND-POS} and, thus, \eqref{2po} is indeed a projection operator.

The inverse transformation \eqref{backtrans} for the density matrix $\rho_{jm}=\mathrm{tr}_{B}\left(\Pi_{jm}\rho\right)$ is defined by
\begin{equation}
\rho_{jm}^2=e^{iA m S_z}\rho_{jm}^2 e^{-iA m S_z}.
\end{equation}
The above transformation gives the identity for the diagonal elements.

The Nakajima-Zwanzig master equation \eqref{NJgen} for the projector \eqref{1po} and the Hamiltonian \eqref{intham} has the form
\begin{eqnarray}\label{me1}
\dot{\rho}^1_{S}(t)=-\frac{A^2}{4}\int_0^t dt_1 \left\lbrace \Omega_+ S_+e^{-i A S_z(t-t_1)}S_-\rho_{S}^1(t_1)\right. \\
-\Omega_-S_+e^{-i A t S_z}\rho_{S}^1(t_1)e^{i A t_1 S_z}S_-\nonumber\\
- \Omega_+e^{i A t_1 S_z}S_-\rho_{S}^1(t_1)S_+e^{-i A t S_z}\nonumber\\
+\left.\Omega_-
\rho_{S}^1(t_1)e^{-i A (t-t_1) S_z}S_-S_++\mathrm{H.c.}\right\rbrace\nonumber,
\end{eqnarray}
where the $\Omega_\pm$ is defined as
\begin{eqnarray}
\Omega_\pm&=&N\cos^{N-1}((A(t-t_1)+i\beta)/2)\\
&&\!\!\!\!\!\!\!\!\!\!\!\!\times\exp[-i A(t-t_1)/2]\exp[\pm\beta/2]/(2 \cosh^N(\beta/2)).\nonumber
\end{eqnarray}
The second order time-convolutionless master equation is obtained by replacing $\rho(t_1)$ with $\rho(t)$.

By substituting Eqs.~\eqref{intham} and \eqref{2po}  into Eq.~\eqref{NJgen} one obtains the Nakajima-Zwanzig master equation in the form
\begin{eqnarray}\label{me2}
\dot{\rho}^2_{jm}(t)=-\frac{A^2}{4}\int_0^t dt_1 \left\lbrace \tilde{\Omega}_+^{m} S_+e^{-i A S_z(t-t_1)}S_-\rho_{jm}^2(t_1)\right. \\
-\tilde{\Omega}_-^{m+1}S_+e^{-i A t S_z}\rho_{jm+1}^2(t_1)e^{i A t_1 S_z}S_- \nonumber\\
- \tilde{\Omega}_+^{m-1}e^{i A t_1 S_z}S_-\rho_{jm-1}^2(t_1)S_+e^{-i A t S_z}\nonumber\\
+\left.\tilde{\Omega}_-^{m}
\rho_{jm}^2(t_1)e^{-i A (t-t_1) S_z}S_-S_++\mathrm{H.c.}\right\rbrace \nonumber,
\end{eqnarray}
where 
\begin{eqnarray}\tilde{\Omega}_+^m=(j(j+1)-m(m+1))e^{i A m(t-t_1)},\\
\tilde{\Omega}_-^m=(j(j+1)-m(m-1))e^{i A (m-1)(t-t_1)}.
\end{eqnarray}
The initial conditions for Eq.~\eqref{me2} can be found by substituting  the initial conditions \eqref{initial} into Eq.~\eqref{2po}. Explicitly, one gets
\begin{equation}\label{init}
\rho_{jm}^2(0)=\frac{e^{-\beta m }  N_j}{z}\rho_S(0).
\end{equation}
Of course, the time-convolutionless master equation is derived from \eqref{me2} by replacing $\rho(t_1)$ with $\rho(t)$.

The relevant part of the density matrix is defined from \eqref{me2} as
\begin{equation}\label{tot}
\rho_S^2(t)=\sum\limits_{j=0,\frac{1}{2}}^{N/2}\sum\limits_{m=-j}^{j}\rho_{jm}^2(t).
\end{equation}

\subsection{Example $j=1$}

In this section we apply the general theory derived in the previous sections to a concrete example, namely, to a central spin $j=1$ coupled to a spin bath. The master equation \eqref{me1} can be solved numerically (for both  Nakajima-Zwanzig and TCL equations). The dynamics of state $|1,1\rangle$ for spin $j=1$ is shown in Figs.~4-6 and 9. We notice, that for this type of projectors the TCL master equation gives the better result. Nonetheless, change of the inverse temperature has no effect on the accuracy of the methods (compare Figs. 4 and 5). The increase of the number of spins into the bath leads to a decrease of accuracy (compare Figs. 4 and 6). So, we can conclude that the accuracy of the TCL master equation is $\sim 2/(A\sqrt{N})$ and of the Nakajima-Zwanzig master equation $\sim 1/(A\sqrt{N}).$

\begin{figure}
\includegraphics[scale=0.5]{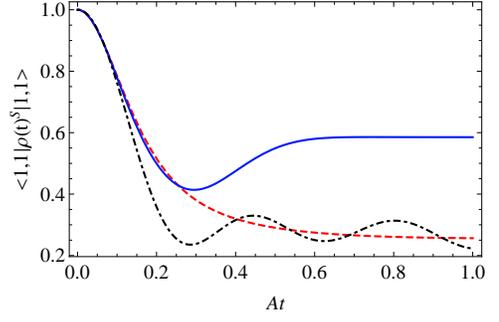}
\caption{(Color online) The comparison of the exact and approximate dynamics given by \eqref{me1} of the upper state for the spin $j=1$ in a spin bath. The blue solid curve is the exact solution, the red dashed curve is the TCL solution and the black dot-dashed curve is the Nakajima-Zwanzig solution. The inverse  temperature of the bath is $\beta=0.25$. The number of spins in the bath is $N=101$. The initial state is $\rho_S(0)=|1,1\rangle\langle1,1|$. The coupling constant is $A=1.$}\label{fig3}
\end{figure}

\begin{figure}
\includegraphics[scale=0.5]{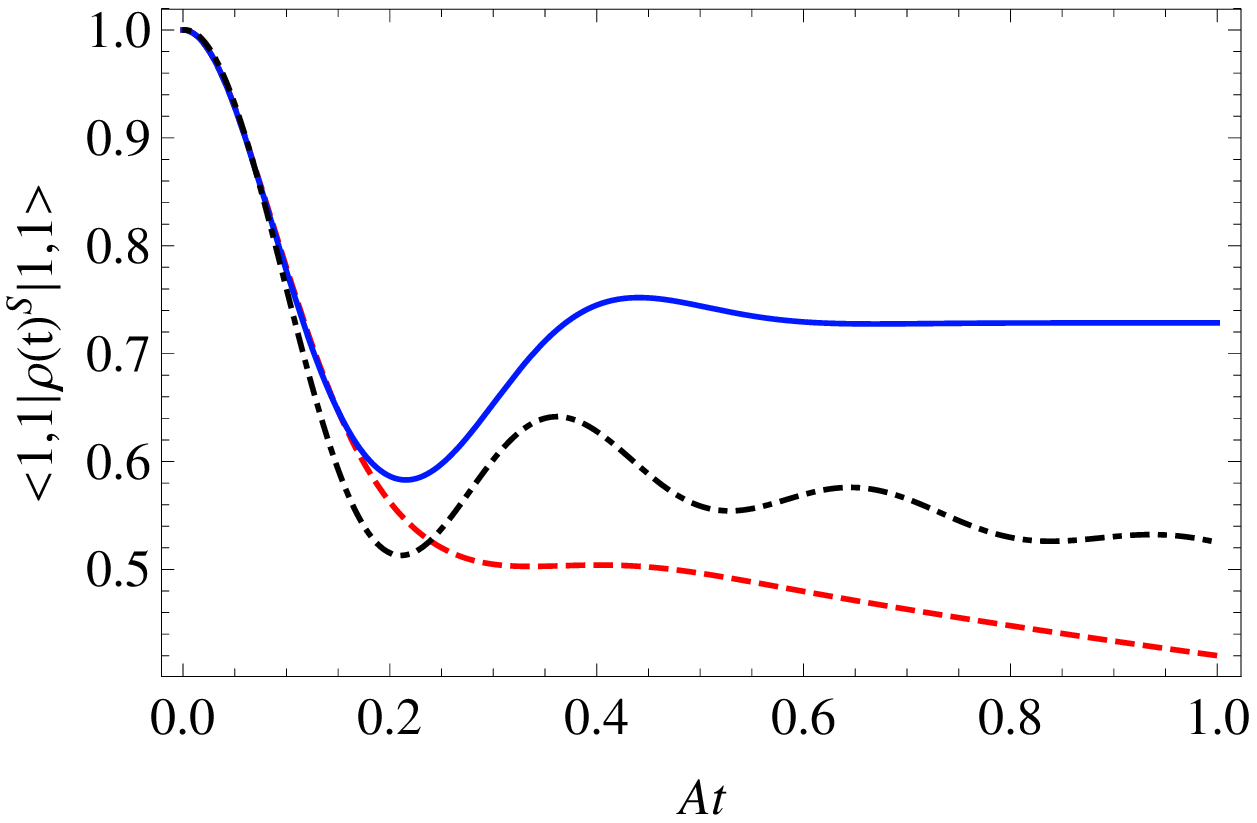}
\caption{(Color online) The comparison of the exact and approximate dynamics given by \eqref{me1} of the upper state for the spin $j=1$ in a spin bath. The blue solid curve is the exact solution, the red  dashed curve is the TCL solution and the black dot-dashed curve is the Nakajima-Zwanzig solution. The inverse temperature of the bath is $\beta=0.5$. The number of spins in the bath is $N=101$. The initial state is $\rho_S(0)=|1,1\rangle\langle1,1|$. The coupling constant is $A=1.$}\label{fig4}
\end{figure}

\begin{figure}
\includegraphics[scale=0.5]{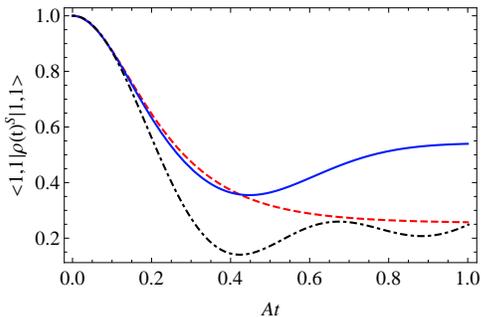}
\caption{(Color online) The comparison of the exact and approximate dynamics given by \eqref{me1} of the upper state for the spin $j=1$ in a spin bath. The blue solid curve is the exact solution, the red dashed curve is the TCL solution and the black dot-dashed curve is the Nakajima-Zwanzig solution. The inverse temperature of the bath is $\beta=0.25$. The number of spins in the bath is $N=51$. The initial state is $\rho_S(0)=|1,1\rangle\langle1,1|$. The coupling constant is $A=1.$}\label{fig5}
\end{figure}

The master equation \eqref{me2} is more complicated than  the corresponding Eq.~\eqref{me1} and we discuss it in detail. As one can see the Eq.~\eqref{me2} has a non-trivial dependence on the index $m$. The index $m$ can take on a value $-j,...,j$ and consequently the system consists of independent blocks of $k^2\times(2j+1)$  equations (we assumed $\rho_{j,|j|+1}=0$), where $k^2$ is the dimension of the density matrix of the central spin. Nonetheless, it can be easily shown that the diagonal elements of any density matrix $\rho_{jm}$ are independent of the off-diagonal elements. 

The structure of the projection operator \eqref{2po} allows to extract from Eq.~\eqref{me2} a minimal system of $(k-1)$ coupled equations. Actually, the number of the coupling equation for the diagonal elements is $k$, but the conservation of the total spin of the system gives one more relationship between the equations.

Now, we are ready to build the system of equations defining the master equation \eqref{me1} for the case of a spin equal to 1. The dimension of the relevant density matrix in this case is $(3\times 3)$, so the system reads

\begin{eqnarray}
\dot{\rho}^{11}_{jm-1}(t)&=&A^2/4 \int\limits_{0}^{t}ds \cos(A(m-1)(t-s))\label{TCL1} \\ &&\!\!\!\!\!\!\!\!\!\!\!\!\!\!\!\!\!\!\!\!\!\!\!\!\!\!\times(j-m+1)(j+m)\left(C-2\rho^{11}_{jm-1}(s)-\rho^{33}_{jm+1}(s)\right),\nonumber\\ 
\dot{\rho}^{33}_{jm+1}(t)&=&A^2/4 \int\limits_{0}^{t}ds \cos(A(m+1)(t-s))\label{TCL2} \\ &&\!\!\!\!\!\!\!\!\!\!\!\!\!\!\!\!\!\!\!\!\!\!\!\!\!\!\times(j+m+1)(j-m)\left(C-\rho^{11}_{jm-1}(s)-2\rho^{33}_{jm+1}(s)\right).\nonumber   
\end{eqnarray}
In the above equations we introduced the constant $C=\rho^{11}_{jm-1}(t)+\rho^{22}_{jm}(t)+\rho^{33}_{jm+1}(t)=\rho^{11}_{jm-1}(0)+\rho^{22}_{jm}(0)+\rho^{33}_{jm+1}(0)$. The initial condition is given by Eq.~\eqref{init}.

The above integro-differential equation can be solved analytically with help of the Laplace transformation. In particular, the $\rho^{11}_{jm-1}(t)$ for the initial state $\rho_S(0)=|1,1\rangle\langle 1,1|$ is
\begin{widetext}
\begin{equation}\label{NZ2}
\rho^{11}_{jm-1}(t)=\frac{C \left(2 K \left(j^2+j-m+1\right)+(j-m+1) (j+m) ((K-J) \cosh (t {X_-})+(J+K) \cosh (t {X_+}))-2 J^2\right)}{2 K^2-2 J^2},
\end{equation}
\end{widetext}
where $$J=\sqrt{(j-m) (j-m+1) (j+m) (j+m+1)},$$  $$K=2 j (j+1)-m^2+1,$$ $$X_\pm=A/4\sqrt{-K\pm J}.$$ 
The characteristics of the spin can be found from the above solution using \eqref{tot}.

\begin{figure}
\includegraphics[scale=0.5]{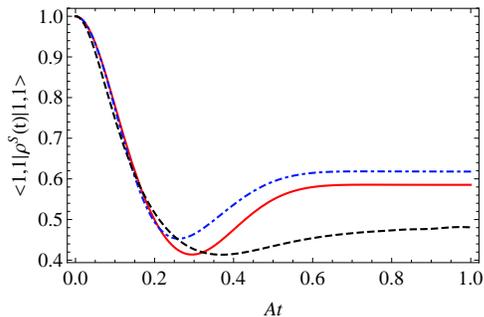}
\caption{(Color online) The comparison of the exact and approximate dynamics given by \eqref{Tsol}, \eqref{Tso2} and \eqref{NZ2} of the upper state for the spin $j=1$ in a spin bath. The red solid curve is the exact solution, the black dashed curve is the TCL solution and the blue dot-dashed curve is the Nakajima-Zwanzig solution. The inverse temperature of the bath is $\beta=0.25$. The number of spins in the bath is $N=101$. The initial state is $\rho_S(0)=|1,1\rangle\langle1,1|$. The coupling constant is $A=1.$}\label{fig6}
\end{figure}

\begin{figure}
\includegraphics[scale=0.5]{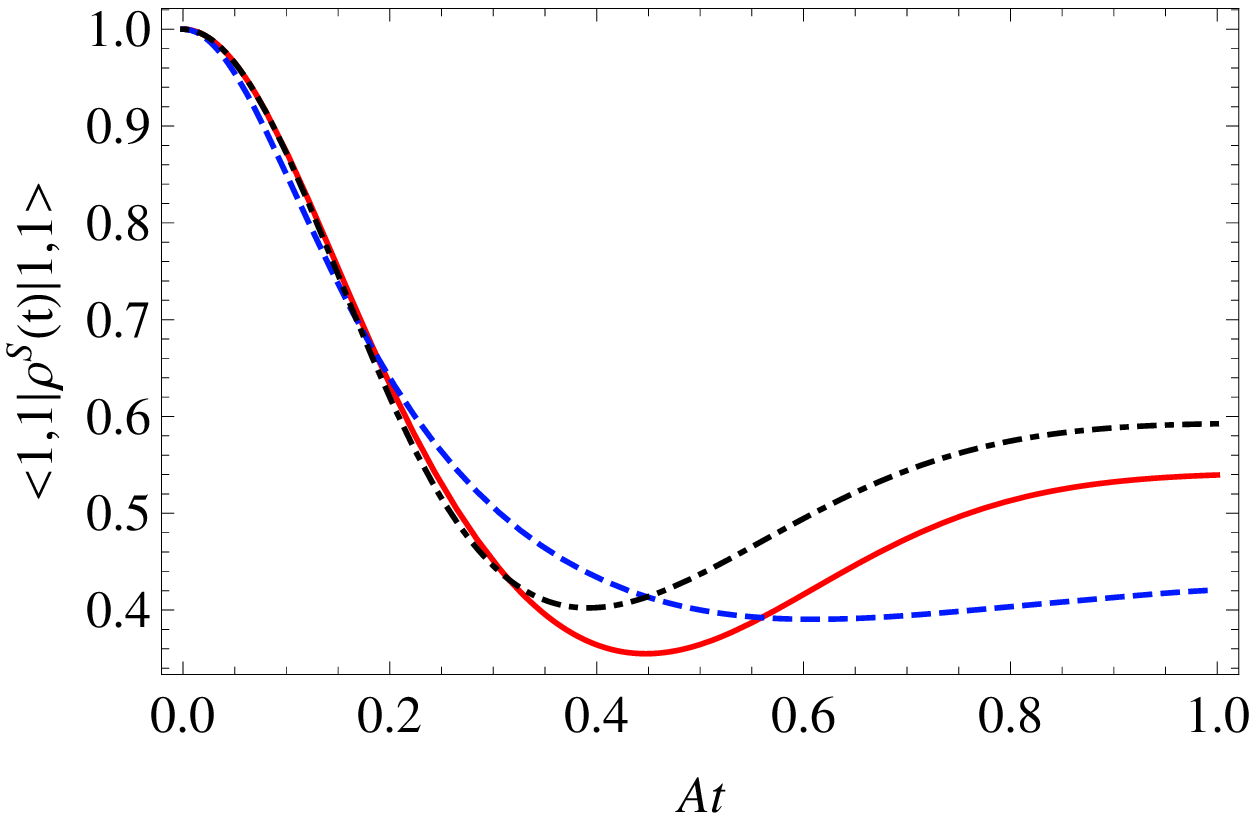}
\caption{(Color online) The comparison of the exact and approximate dynamics given by \eqref{Tsol}, \eqref{Tso2} and \eqref{NZ2} of the upper state for the spin $j=1$ in a spin bath. The red solid curve is the exact solution, the blue dashed curve is the TCL solution and the black dot-dashed curve is the Nakajima-Zwanzig solution. The inverse  temperature of the bath is $\beta=0.25$. The number of spins in the bath is $N=51$. The initial state is $\rho_S(0)=|1,1\rangle\langle1,1|$.  The coupling constant is $A=1.$}\label{fig8}
\end{figure}

\begin{figure}
\includegraphics[scale=0.5]{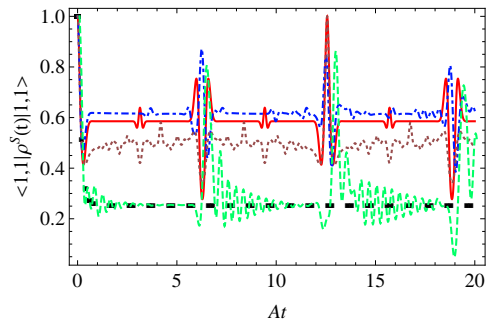}
\caption{(Color online) The comparison of the exact and approximate long-time dynamics of the upper state for the spin $j=1$ in a spin bath. The red solid curve is the exact solution, the green dashed curve is the Nakajima-Zwanzig \eqref{me1} solution, the blue dot-dashed curve is the Nakajima-Zwanzig solution \eqref{NZ2}, the brown dotted curve is TCL  \eqref{Tsol}-\eqref{Tso2} and the black symbols of the squares are the TCL \eqref{me1} solution. The inverse  temperature of the bath is $\beta=0.25$. The number of spins in the bath is $N=101$. The initial state is $\rho_S(0)=|1,1\rangle\langle1,1|$. The coupling constant is $A=1.$}\label{fig9}
\end{figure}

The TCL master equation has the same form of the corresponding Nakajima-Zwanzig master equation \eqref{TCL1}-\eqref{TCL2}. Namely,  the diagonal matrix elements is governed by the system 
\begin{eqnarray}
\dot{\rho}^{11}_{jm-1}(t)&=&A^2/4 \sin(A(m-1)t)/(m-1)\label{TCL3} \\ &&\!\!\!\!\!\!\!\!\!\!\!\!\!\!\!\!\!\!\!\!\!\times(j-m+1)(j+m)\left(C-2\rho^{11}_{jm-1}(t)-\rho^{33}_{jm+1}(t)\right),\nonumber\\ 
\dot{\rho}^{33}_{jm+1}(t)&=&A^2/4  \sin(A(m+1)t)/(m+1)\label{TCL4} \\ &&\!\!\!\!\!\!\!\!\!\!\!\!\!\!\!\!\!\!\!\!\!\times(j+m+1)(j-m)\left(C-\rho^{11}_{jm-1}(t)-2\rho^{33}_{jm+1}(t)\right).\nonumber   
\end{eqnarray}
The analytical solution of the above system of equations can be found as a series. Unfortunately,  the series converges extremely slowly. Moreover, the system  is not stiff and methods of numerical solution diverge. 

Nevertheless, we can find an approximate analytical solution for the case $m\gg 1$. The solution of the system in this case is
\begin{eqnarray}
\rho^{11}_{jm}(t)=\frac{2C+2X+Y}{6},\label{Tsol}\\
\rho^{33}_{jm}(t)=\frac{2C-X-2Y}{6}.\label{Tso2}
\end{eqnarray}
In the above equation 
$$X=\exp\{\frac{A\left(j^2-m^2\right) (\cos (m t)-1)}{4m^2}\},$$ 
$$Y=\exp\{\frac{3A \left(j^2-m^2\right) (\cos (m t)-1)}{4m^2}\}.$$ 
The expressions \eqref{Tsol} -\eqref{Tso2}
are a good approximation for the case $m \gg 1$ and, because the  required expression is given by \eqref{tot} and every term given by \eqref{Tsol} -\eqref{Tso2} is small we can conclude that such approximation gives a good result for a large number of spins in the bath.

The solution Eq.~ \eqref{me2} for the central spin equal to 1 is presented in Figs.~7-9. One can see, that the Nakajima-Zwanzig master equation \eqref{NZ2} leads to results similar to \eqref{Tsol}-\eqref{Tso2} for the short time $\sim 2/(A\sqrt{N}$). Nevertheless, the solution \eqref{NZ2} gives a better result than the approximate solution of the corresponding TCL master equation \eqref{Tsol}-\eqref{Tso2}. Thus, the method of correlated projection operators gives the same result as the previous one for short time-scales $~\sim 2/(A\sqrt{N})$.

Another picture emerges for the long-time dynamics (Fig.~9). In this domain the projectors \eqref{2po} give the correct qualitative behaviour and allow to predict the increase of some peaks. So, for the description of the long-time dynamics Eq.~\eqref{2po} is more useful.

The Nakajima-Zwanzig equation \eqref{me2} for the  model under consideration is solved analytically, but the corresponding TCL master equation can be solved only approximately. So, for the case of the central spin equal to 1, the use of the Nakajima-Zwanzig master equation for the projectors \eqref{2po} is to be preferred, because it can be solved exactly.

\section{CONCLUSIONS}

In the present article we have investigated the model of an arbitrary spin coupled with a  bath of 1/2-spins.
The exact solution for the reduced density matrix, corresponding to the central spin, was found and used to test a few approximation techniques.

The approximation techniques we have employed are based on time-independent projection operators methods. We have considered two types of projection operators from the general class of the time-independent projectors \eqref{PROJECTION-GENFORM}. Namely, the first operator projects on the equilibrium bath state at a temperature $1/\beta$ and the second operator projects on the eigenspace of the  bath operators $J_z$ and $J^2$. For both choices of projection operators we have derived the second order Nakajima-Zwanzig master equations \eqref{me1} and \eqref{me2}, and also the corresponding TCL master equations. Every master equation was applied to a model of  central spin equal to one.

As the main result of this article we indicate that the traditional way to approximate open spin systems based on the time-independent projection operators gives the correct short-times dynamics of the relevant part of the system. The time-scales of the applicability of such approximate method can be estimated 
of the order of $1/(A\sqrt{N}),$ where $A$ is the coupled constant and $N$ is the number of spins in the bath. It seems, that approximation methods based on time-independent projection techniques are not so accurate for a large numbers of spins in the bath. The dependence on $N$ can be removed by renormalization  of the interaction constant, as $A=A_\mathrm{eff}/\sqrt{N}$. In this case  the time-scale for the applicability of the second order approximation will scale as $~1/A_\mathrm{eff}$. Some authors include such a factor in the initial Hamiltonian \cite{SPIN,4}. In this sense, we have formally derived the meaning of this factor.

For spin star models the accuracy substantially depends on the number of spins in a spin bath. The time-scales of accurate reproduction of the exact dynamics can be estimated as $~\sim 1/(A\sqrt{N})$. Nevertheless, the TCL master equation with the traditional projectors \eqref{1po} gives better result than the corresponding Nakajima-Zwanzig master equation, and, moreover, gives similar results as the correlated projection operator  \eqref{2po} for short time-scales dynamics. In order to predict short time-scales dynamics it is more useful to use  the simplest projection operators combined with the technique of the TCL master equation.

In this connection we notice that the correlated projection operator technique may be useful for the prediction of the long-time behaviour of the spin system, if the analytical solution of the correlated projection operator master equation can be found. However, we emphasize that this result is beyond the scope of applicability of the second-order approximations and is rather a nice exception. In every concrete case long-time dynamics must be investigated additionally with the help of other methods.    

Note that all the approximate equations of the second order presented in the paper give an equivalent dynamics, and can be easily extended to the non-resonant case, as shown in the appendix of this paper. Thus, it is advisable to use the simplest approximate equation, namely, the TCL master equation derived with help of the traditional projection operator. It might be helpful to note, all approximate techniques have limits of applicability 
and here  we have shown \ 
  the limit of applicability of the projection methods of the second-order. Finally,  it seems that  the traditional methods do not work well for spin bath models. Thus, it is necessary to develop new approximation techniques for strongly correlated open systems.

\begin{acknowledgments}
This work is based upon research supported by the South African
Research Chair Initiative of the Department of Science and
Technology and National Research Foundation.
\end{acknowledgments}

\appendix
\section{Non-resonant case}

 In this appendix we consider the non-resonant case of the simplest model of the 1/2-spin coupled to a bath of spins \cite{CRL} and show that the correlated projection operator technique does not give appreciable advantages over the traditional one. 

Let us consider the non-resonant problem of an 1/2-spin coupled to a spin bath in a spin star configuration. This simplest case  allows to find an analytical solution of the non-resonant problem \cite{CRL}. The Hamiltonian for the system has the form
\begin{equation}
H=\omega_0 S_z+A\mathbf{S J}, 
\end{equation}  
which differs from  \eqref{hen_ham} only in the free term. In the above expression the $\omega_0$ is the frequency detuning. As it was noticed in \cite{CRL}, this problem has an analytic solution, due to the fact that the two-dimensional subspaces spanned by the states $|+\rangle\otimes|j,m\rangle$ and $|-\rangle\otimes|j,m+1\rangle$ are invariant under the time evolution. For simplicity we consider the initial conditions \eqref{initial} with $\beta=0.$ 

The exact elements of the reduced density matrix \eqref{reduced} can be written as \cite{diff}

\begin{eqnarray}
\rho_{11}^S(t)\!\!\!\!&=&\!\!\!\!\sum_{jm}\left\lbrace\! \left[\cos^2(\mu_+(j,m)t)+\frac{K_+^2(m)}{4\mu_+^2(j,m)}\sin^2(\mu_+(j,m)t) \right]\right.\nonumber \\
&&\!\!\!\!\!\!\!\!\!\!\!\!\!\!\!\!\!\!\!\!\times\rho_{11}^S(0)+\left. \left[\frac{A^2b^2(j,m)}{4\mu_+^2(j,m)}\sin^2(\mu_+(j,m)t)\rho_{22}^S(0)  \right]\right\rbrace\frac{N_j}{2^N},\\
\rho_{12}^S(t)\!\!\!\!&=&\!\!\!\!\sum_{jm}\left[\cos(\mu_+(j,m)t)-\frac{iK_+(m)}{2\mu_+(j,m)}\sin(\mu_+(j,m)t) \right]\nonumber\\
&&\!\!\!\!\!\!\!\!\!\!\!\!\!\!\!\!\!\!\!\!\!\!\!\times\!\!\left[\cos(\mu_-(j,m)t)\!+\!\frac{iK_-(m)}{2\mu_-(j,m)}\sin(\mu_-(j,m)t) \right]\!\!\!\frac{\rho_{12}^S(0)N_j}{2^N}\!,   
\end{eqnarray}
where $\mu_\pm(j,m)=\sqrt{K_\pm^2(m)+A^2b^2(j,\pm m)}/2,$ $K_\pm(m)=\pm\omega_0+A(\pm m+1/2)$ and $b(j,m)=\sqrt{(j-m)(j+m+1)}.$ The Eqs.~(A2) and (A3)  are valid for any initial conditions of the form \eqref{initial}.

The interaction Hamiltonian for the considering problem can be written as
\begin{eqnarray}\label{intham1}
H_I=\frac{A}{2}\left(e^{i\omega_0 t}S_+e^{-i A(S_z-J_z)t}J_-+\mathrm{H.c.}\right).
\end{eqnarray}
Note that the Hamiltonian ~\eqref{intham1} retains its form for any value of the central spin.

The second order master equations for the studied model have forms, which are analogous to \eqref{me1} and \eqref{me2} if we replace the coefficients  $\Omega_\pm$ by $\Omega_\pm e^{i\omega_0(t-t_1)}$ and $\tilde{\Omega}_\pm^m$ by $\tilde{\Omega}_\pm^m e^{i\omega_0(t-t_1)}$. Again, these master equations retain them forms for any value of the central spin.

Since the Nakajima-Zwanzig master equations have no advantage over the TCL ones, we consider only the TCL master equations. The simplest case of the central spin equal to 1/2 allows to build an analytical solution for both types of  the TCL master equations, corresponding to the traditional and the correlated projection operators, and to compare their with the exact one.

\begin{figure}
\includegraphics[scale=1]{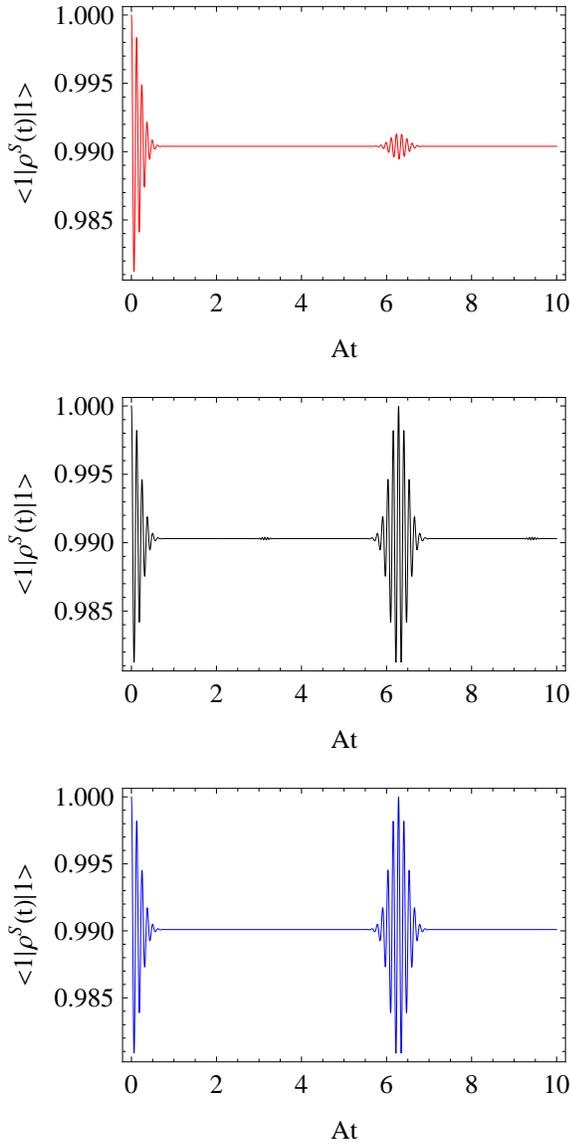}
\caption{(Color online) The non-resonant dynamics of the diagonal element of the density matrix of the central spin equal to 1/2. The red curve (top) is the exact solution, the blue curve (top) is the approximation solution \eqref{tcl1}, the black curve (medium) is the approximation solution \eqref{tcl3}. The number of spins in the bath is $N=101$. The initial state is $<1|\rho^S(0)|1>=1$. The detuning is $\omega_0=51A$  The coupling constant is $A=1.$}\label{fig10}
\end{figure}

\begin{figure}
\includegraphics[scale=1]{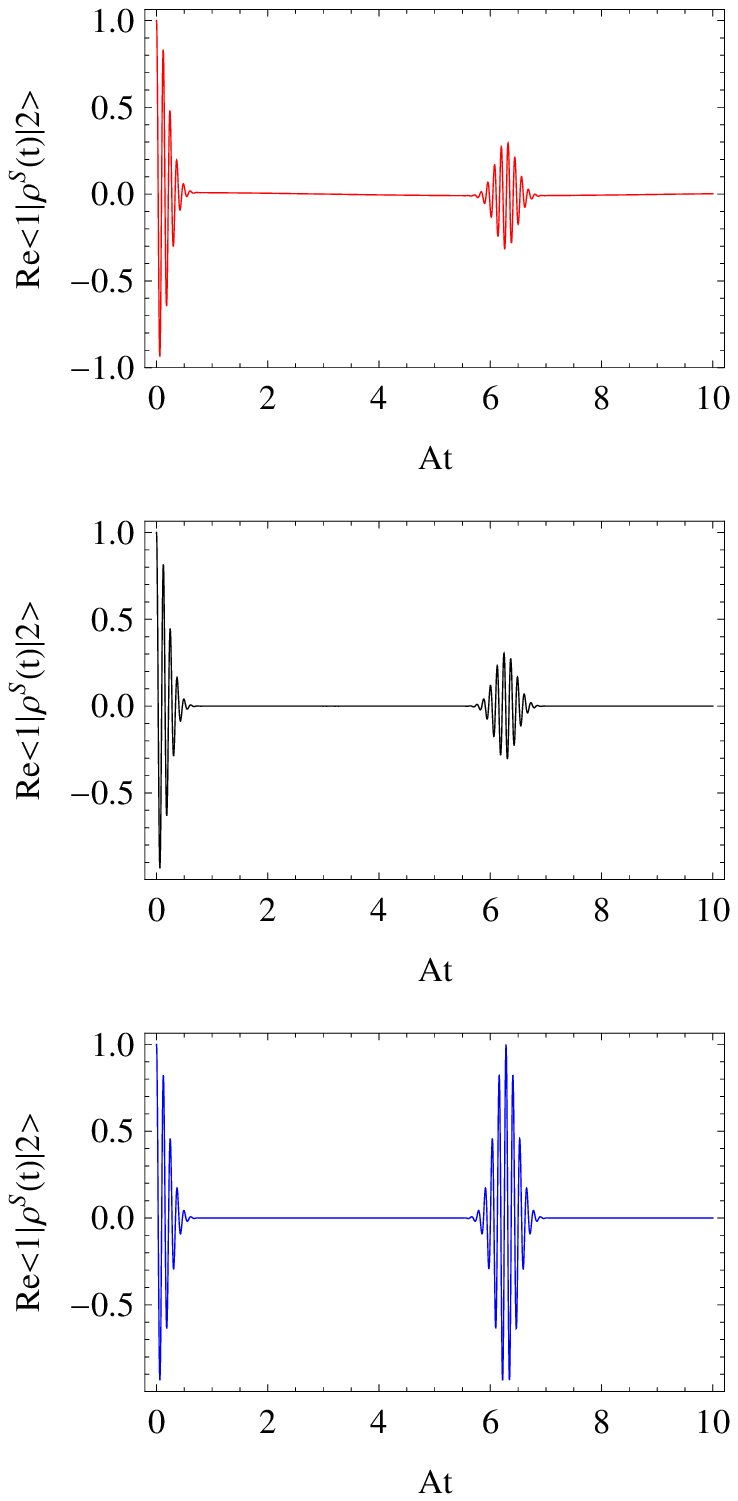}
 \caption{(Color online) The non-resonant dynamics of the off-diagonal element of the density matrix of the central spin equal to 1/2. The red curve (top) is the exact solution, the blue curve (bottom) is the approximation solution \eqref{tcl2}, the black curve (medium) is the approximation solution \eqref{tcl4}. The number of spins in the bath is $N=101$. The initial state is $<1|\rho^S(0)|2>=1$. The detuning is $\omega_0=51A.$  The coupling constant is $A=1.$}\label{fig11}
\end{figure}

The solutions of the TCL master equations \eqref{me1} and \eqref{me2} in this case is written as

\begin{eqnarray}
\langle 1|\rho^S_1(t)|1\rangle&=&\frac{1}{2}+\left( \langle 1|\rho^S_1(0)|1\rangle-\frac{1}{2}\right) \label{tcl1}\\
&&\quad\quad\times\exp[-\frac{A^2}{2}\int\limits_{0}^{t}f(t_1)dt_1],\nonumber\\
\langle 1|\rho^S_1(t)|2\rangle\!\!\!\!&=&\!\!\langle 1|\rho^S_1(0)|2\rangle \exp[-\frac{A^2}{4}\!\!\int\limits_{0}^{t}\!\!C(t_1)dt_1],\label{tcl2}\\
\langle 1|\rho^S_{2jm}(t)|1\rangle&\!\!\!=\!&\!\!\frac{N_j}{2^{N+1}}+\left(\!\langle 1|\rho^S_{2jm}(0)|1\rangle-\frac{N_j}{2^{N+1}}\!\right)\label{tcl3} \\ &&\quad\quad\times\exp[-A^2\int\limits_{0}^{t}B(t_1)dt_1],\nonumber\\
\langle 1|\rho^S_{2jm}(t)|2\rangle\!\!\!&=&\!\!\!\langle 1|\rho^S_{2jm}(0)|2\rangle\! \exp[-\!\frac{A^2}{4}\!\!\int\limits_{0}^{t}\!\!D(t_1)dt_1]\label{tcl4}.
\end{eqnarray} 

In the above expressions we introduced the following functions
\begin{eqnarray}
f(t)&=&C(t)+C^*(t),\nonumber\\
C(t)&=&\frac{1}{2}\int\limits_{0}^{t}dt_1N\cos^{N-1}(A(t-t_1)/2)\exp(i\omega_0(t-t_1)),\nonumber\\
B(t)&=&\int\limits_{0}^{t}dt_1b(j,m)\cos(K_+(m)(t-t_1)),\nonumber\\
D(t)&=&\int\limits_{0}^{t}dt_1[b(j,-m)\exp(-i K_-(t-t_1))\nonumber\\&+&b(j,m)\exp(i K_+(t-t_1))].\nonumber
\end{eqnarray} 
All the integrals appearing in the expression can be easily calculated. We do not explicitly show them as the expressions are rather cumbersome.

Now, we have to make a transformation back to the Schr\"odinger picture to be able to compare the exact and the approximation solutions. It is clear that the diagonal elements of the density matrix do not change under the transformation. The off-diagonal elements change to 
\begin{eqnarray}
&\langle&\!\!\! 1|\rho^S_{2jm}(t)|2\rangle_{SP}\!\!=\!\!\exp[i(\omega_0+Am) t]\langle 1|\rho^S_{2jm}(t)|2\rangle\!,\\
&\langle&\!\!\! 1|\rho^S_{1}(t)|2\rangle_{SP}\!\!=\!\!\exp[i\omega_0 t]\cos^N[At/2]\langle 1|\rho^S_{1}(t)|2\rangle\!.
\end{eqnarray}
where the subscript ``SP'' means ``Schr\"odinger picture''. 

The comparison of both the approximation solutions with the exact one in the off-resonant case is shown in Figs.~10-11. One can see that the diagonal element of the density matrix (Fig.~10) describes well both types of the approximation equations. Moreover, it can be shown that  both approximation solutions give the same behaviour in the long time-scales. So, the diagonal matrix elements are described with the same accuracy  both approximated solutions (A5)-(A8).

The dynamics of the off-diagonal elements of the density matrix is shown in Fig.~11. It seems that in this case the correlated projection operator works better. The solution following from  the correlated projector master equation describes the amplitude of oscillations more precisely, but the oscillations have a different phase and  do not allow to predict the exact dynamics with reasonable accuracy, only the qualitative behaviour. Thus, the applicability of the second order approximate methods can be estimated as $1/(\omega_0+A)\sqrt{N},$ where $\omega_0$ is the detuning.

Notice, that the dynamics of the off-diagonal elements of the density matrix (Eqs.~(A6) and (A8)) presented in \cite{CRL} seems to have another behaviour. 
It is due to the authors \cite{CRL} represent them results in the interaction picture and we represent our results in the Schr\"odinger picture.


Let us consider the  differences between \eqref{tcl1}-\eqref{tcl4} and the results derived in \cite{CRL}. The correlated projection operator master equation \eqref{me2} gives a dynamics which completely agrees with the corresponding results in  \cite{CRL} (see \cite{diff}). Nevertheless, the results derived with the help of the traditional master equation differ from the corresponding results in \cite{CRL} (see Eq.~(58) therein). The results corresponding to Eq.~(58 in \cite{CRL}) hold only for $\cos(At/2)=1$ under the integral in Eq.~\eqref{tcl1}. This is true only if $NA\ll\omega_0,$ which corresponds to the absence of the system-bath interaction.



\begin{thebibliography}{10}
\bibitem{1} R.Lo Franco, B.Bellomo, S.Maniscalco, and G.Compagno, Int. J. Mod. Phys. B \textbf{27}, 1345053 (2013)


\bibitem{SPIN} N.V. Prokof'ev and P.C.E. Stamp, Rep. Prog. Phys. \textbf{63}, 669 (2000).

\bibitem{itqss} J.B. Parkinson, D.J.J. Farnell, \textit{An Introduction to Quantum Spin Systems,} Lect. Notes
Phys. 816 (Springer, Berlin Heidelberg, 2010).

\bibitem{glas} T.F. Rosenbaum,   J. Phys. Chem. \textbf{8}, 9759 (1996)

\bibitem{cond} A. J. Leggett et al,  Rev. Mod. Phys. \textbf{59}, 1 (1987)

\bibitem{cond2} A. O. Caldeira  and A. J. Leggett   Ann. Phys., NY \textbf{149}, 374 (1983)

\bibitem{super} A. L. Efros  and M. Pollak (ed) \textit{ Electron–electron Interactions in Disordered Systems} (Amsterdam: NorthHolland, 1985)

\bibitem{toqs} H.-P. Breuer and F. Petruccione, \textit{The Theory of Open Quantum Systems} (Oxford University Press, 2002).

\bibitem{NAKAJIMA} S. Nakajima, Prog. Theor. Phys. \textbf{20}, 948 (1958).

\bibitem{ZWANZIG} R. Zwanzig, J. Chem. Phys. \textbf{33}, 1338   (1960).  


\bibitem{TCL_first} S. Chaturvedi and F. Shibata, Z. Physik B \textbf{35}, 297 (1979).

\bibitem{oldTCL} H.-P. Breuer, D. Burgarth, F. Petruccione, Phys. Rev. B \textbf{70}, 045323 (2004).


\bibitem{CRL}J. Fischer, H.-P. Breuer, Phys. Rev. A \textbf{76}, 052119 (2007).

\bibitem{Repke} D. Zubarev, V. Morozov and G.R\"{o}pke \textit{Statistical Mechanics of Non-equilibrium Processes. Vol. 1} (Akademie Verlag, 1996).


\bibitem{4} Y. Hamdouni and F. Petruccione, Phys. Rev. B \textbf{76}, 174306 (2007).

\bibitem{fannes} Y. Hamdouni, M. Fannes, and F. Petruccione, Phys. Rev. B \textbf{73}, 245323 (2006).

\bibitem{Varsh} D.A. Varshalovich, A.N. Moskalev, V.K. Khersonskii \textit{Quantum Theory of Angular Momentum} (World Scientific, 1988).


\bibitem{CRL0} H.-P. Breuer, Phys. Rev. A \textbf{75}, 022103 (2007).

\bibitem{CRL2} H.-P. Breuer, J. Gemmer, M. Michel, Phys. Rev. E \textbf{73}, 016139 (2006).


\bibitem{my} V. Semin, I. Sinayskiy and F. Petruccione, Phys. Rev. A \textbf{86}, 062114 (2012).

\bibitem{manz} A.V. Mangirov, A.D. Polyanin, \textit{Handbook of Integral Equations} (Factorial Press, Moscow, 2000).

\bibitem{diff} Note, that we use the spin matrices, which differ from the Pauli matrices by factor 1/2. Thus, our results are consistent with the results of [13] if to make the change of variable $A$ to $A_B/4$, where $A_B$ is the coupling constant in [13].

\end{thebibliography}
\end{document}